\documentclass[]{article}

\usepackage{alanstyle}
\usepackage{graphicx}
\usepackage{empheq}
\usepackage{apacite}
\usepackage{tikz-cd}
\title{US Equity Risk Premiums during the COVID-19 Pandemic}
\author{Alan L. Lewis\footnote{Newport Beach, California, USA; email: alewis@financepress.com}}

\begin{document}

\maketitle

\begin{abstract}
	We study equity risk premiums in the United States during the COVID-19 pandemic. 

\end{abstract}

\section{Introduction and Summary of Results}

\begin{figure}[h] 
	\caption{{\bf{US COVID-19 development through late April 2020}}} 
	\vspace{10pt}
	\begin{center}
		\includegraphics[width=0.7\textwidth]{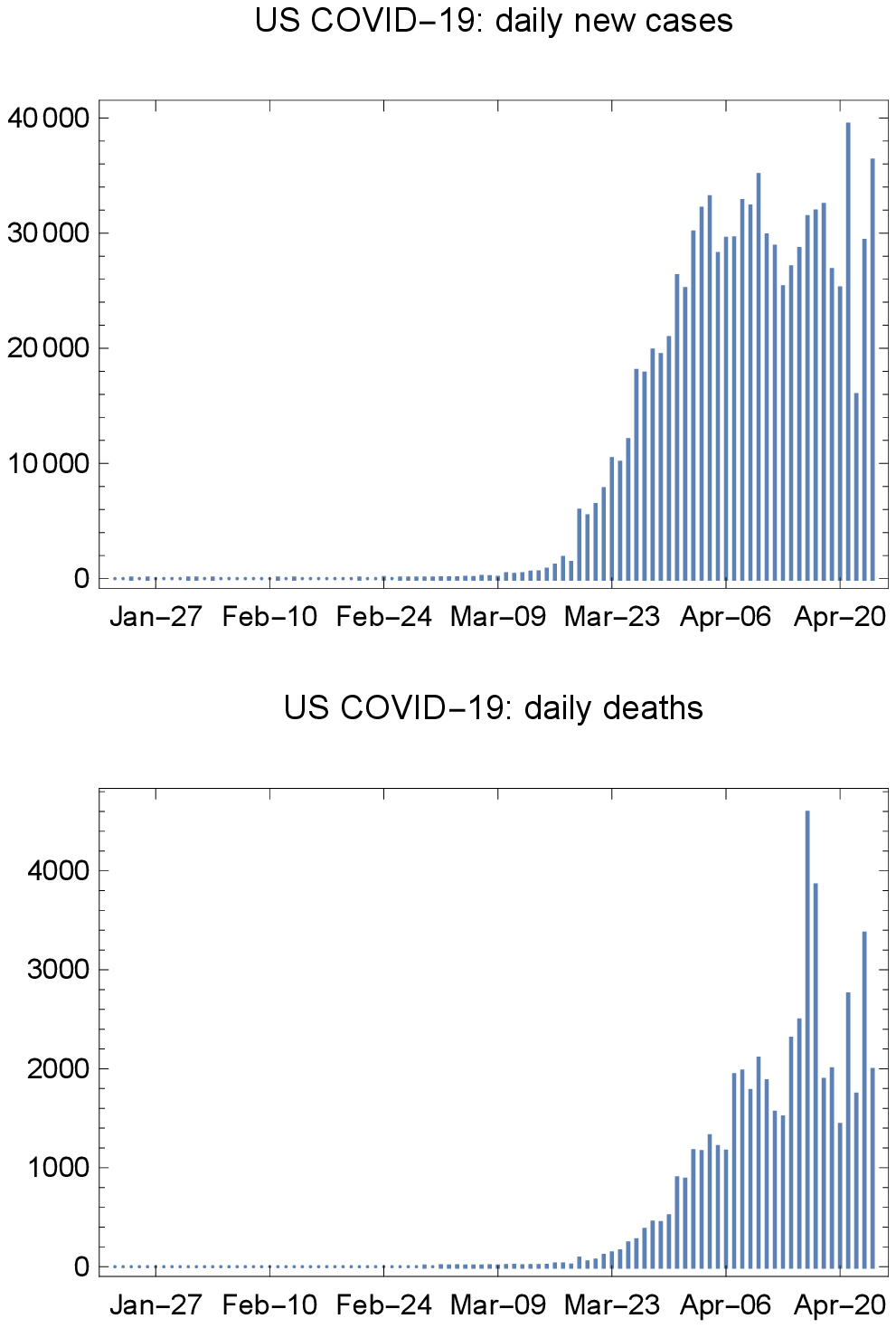}
	\end{center}		
	\label{fig:UScovidplots}
\end{figure}

COVID-19 is shorthand for the novel corona virus disease with origins in Wuhan, China in the fall of 2019. It 
spread in 2020 to become a global pandemic.
As of this writing (late April, 2020), there have been approximately 3 million identified cases worldwide,
and 200,000 deaths. Of those, the US totals are approximately 946,000 cases and 53,000 deaths\footnote{source:
	\url{https://www.worldometers.info/coronavirus/}}. Fig. \ref{fig:UScovidplots} shows the daily US development to date.

Worldwide, governments have urged or mandated shelter-in-place policies,
and mandated shutdowns of most `non-essential' business. In the US this approach has achieved
the immediate goal of buying time for hospitals to prepare for future COVID-19 patients and not be over-whelmed by current ones.

\begin{figure}[h] 
	\caption{{\bf{S\&P500 Index: levels and percent returns}}} 
	\vspace{10pt}
	\begin{center}
		\includegraphics[width=0.9\textwidth]{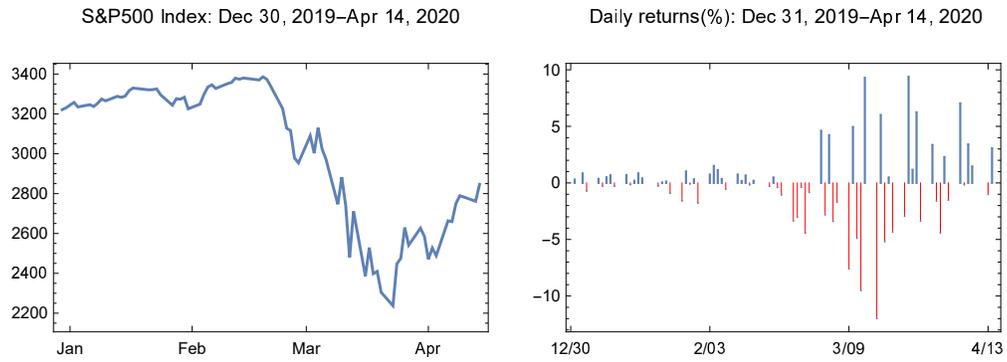}
	\end{center}		
	\label{fig:spxplots}
\end{figure}

But, broad lock-downs are unsustainable for more than a few months. Indeed, many US states are preparing to carefully open up in
mid-May and beyond.  The lock-downs have resulted in enormous economic stress. In the US: 27 million lost jobs and counting at this juncture. These health and economic issues have been a key driver of
recent extreme volatility in financial markets. Fig. \ref{fig:spxplots} shows the S\&P500 index levels and returns during 2020.
For perspective, note that in normal times, a $\pm 3\%$ move would merit a mention and explanation on the nightly news.

Here, we study an interesting and important financial question: what new return expectations have accompanied all
this increased volatility? At first glance, expectations might seem impossible to discern. Indeed,
getting an answer is rather subtle and requires both the options market and estimates of risk aversion. 

Risk averse investors hold equities only if they
feel they will be fairly compensated for the perceived risks. Fair compensation, in the aggregate,
 is called the ``equity risk premium" (ERP). More carefully, the ERP is the market's forward-looking, expected rate of return  -- after
 subtracting an available riskless rate (say a US Treasury rate). Because of the subtraction, it's called an excess return.
 Thus, the ERP can be thought of as the ``required (excess) rate of return", conditioned on what the market knows, to keep all stocks held. Another way to say it: what excess return is needed to clear the equity markets?
 
 Expectations both require a horizon and change with the passage of time: the market learns new things.
 Thus, each day $t$, we have $\mbox{ERP}_{t,T}$ where the $T$ are various time horizons. Our horizons range from one day ahead to just under 3 years. Fixing $t$, a graph of $\mbox{ERP}_{t,T}$ vs. $T$ is an ERP term structure plot.  
 Unlike a familiar interest rate term structure (a yield curve), the ERP term structure is not directly visible 
 and needs to be estimated. Like a yield curve, regardless of the time to the horizon, we always quote ERP's as
  \emph{annual percentage rates}.
  
We study the effect of the pandemic events on the ERP term structures in the United States, from late January through mid-April 2020. Estimates are found using the methods recently developed in \cite{lewis:2019}. We take the
S\&P500 Index as a broad equity market proxy. Then, in brief, daily S\&P 500 index option quotes are combined with estimates of a risk-aversion parameter $\kappa$ to develop the ERP term structures. We briefly review how that works in Sec. \ref{sec:ERPmodel} below. 
Further computational details may be found in Appendix 2 here and the Lewis article.

\pbold{What have I learned?} 
In Sec. \ref{sec:timelines}, one finds the detailed results. My approach is to present brief key-event timelines, show 
corresponding ERP's, and supply some brief commentary. ERP plots   
come with a central estimate (dotted) surrounded by an uncertainty interval in gray. 

Unsurprisingly, there is a strong general association between volatility (VIX levels, for example, as seen in Fig. \ref{fig:vixplots}) and the ERP's. It is well-known that when the market gets very stressed by something, VIX
rises and the whole VIX term structure `inverts'. In other words, short-term (risk-neutral) volatility expectations rise above long-term ones. 
Correspondingly, the ERP term structure also strongly inverts. During the pandemic, this ERP inversion first happened circa Feb 24, 2020; it remained inverted through the end of the study data on Apr 15, 2020. 

\begin{figure}[h] 
	\caption{{\bf{US ERP term structure}}} 
	\vspace{10pt}
	\begin{center}
		\includegraphics[width=0.8\textwidth]{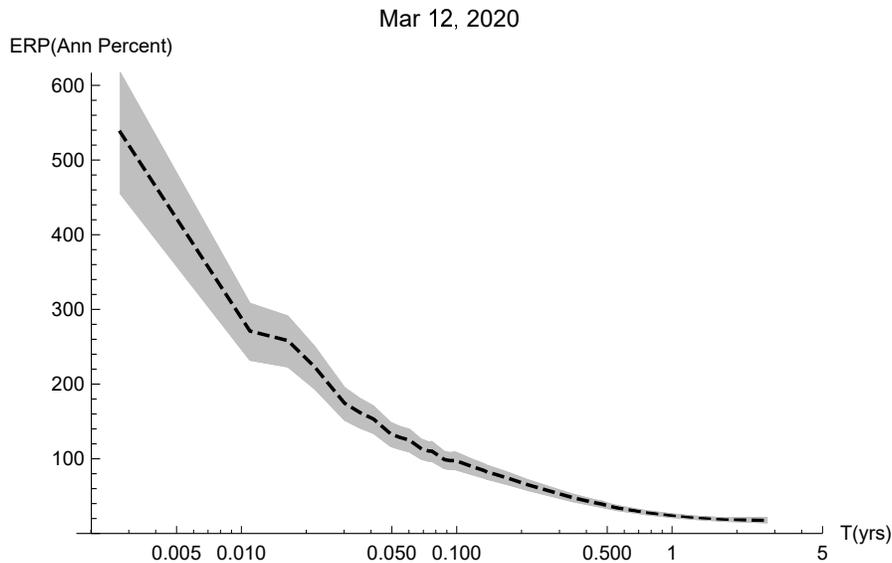}
	\end{center}		
	\label{fig:ERPts031220}
\end{figure}

Qualitatively, based upon my earlier experience with the model in \cite{lewis:2019},
I expected to see these strong inversions. But, I was surprised, quantitatively, by the extraordinary heights reached by the short-term ERP's during mid-March 2020. For example, the March 12 term structure (Fig. \ref{fig:ERPts031220}) shows
the short-dated (one day horizon) ERP reaching a mid-point estimate of approximately 540\% per year! For comparison, long-run 
ERP estimates typically lie in the 3-6\% per year range. Indeed, 3-6\% characterized the pandemic-period US market 
through mid-February. ERP levels as high as the March 12 ones may be record-setting. To say for sure
requires applying the current methodology to option quotes during the 2008-2009 Financial Crisis -- this has not yet been done. 

Finally, some related analysis and updates will be provided in follow-up research to be posted online. See the Outlook at the end for how to locate that.

\section{Brief recap of the ERP model} \label{sec:ERPmodel}

In this section we give a brief, technical explanation of how the ERP's are computed. If you're not interested in these details, feel free to skip ahead to the results in Sec. \ref{sec:timelines}. 

With $\EBB_t$ denoting a (real-world) expectation conditional on  date-$t$ information $I_t$ -- broadly speaking: the ``state of the world" -- we define:

\be   \mbox{ERP}_{t,T}  = \Et{R^e_{t,T}} - R^f_{t,T} = \Et{R^e_{t,T} - R^f_{t,T}}, \quad \mbox{where at time $t$}: \label{eq:ERPdef}  \eb
\begin{itemize}
	\item  $R^e_{t,T}$ is a future random \emph{total} return on the equity market from $t$ to $T$, and
	\item  $R^f_{t,T}$ is a time-$t$ observable risk-free return (using US Treasury instruments).
\end{itemize}
Returns in (\ref{eq:ERPdef}) are simple total returns: $R^e_{t,T} = (\bar{S}_T - \bar{S}_t)/\bar{S}_t$, where $\bar{S}$ is a total-return index incorporating reinvested dividends. (Without a bar, $S_t$ is the price series without dividends). Call $R^e_{t,T} - R^f_{t,T}$ the \emph{excess total return}. 
Like interest rates, we'll always give estimated ERP's on an \emph{annualized percentage basis}. For those, we multiply the ERP calculated from (\ref{eq:ERPdef}) by $100 \times f_{ann}$, where the annualization factor $f_{ann} = 1/(T-t)$, with time measured in years.  

With logarithmic variables $\bar{X}_T  = \log  \bar{S}_T/\bar{S}_t$, and corresponding probability density
$p_{\bar{X}_T}(x)$, (\ref{eq:ERPdef}) is equivalent to

 \be   \mbox{ERP}_{t,T} = \int \e^x p_{\bar{X}_T}(x) \, dx - (1 +  R^f_{t,T}).   \label{eq:ERPdef2}   \eb
 How do we find $p_{\bar{X}_T}(x)$? It turns out that, from the options market (specifically, options on the SPX index),
 we can estimate a closely related probability density $q_{\bar{X}_T}(x)$, the so-called ``risk-neutral" density.
 A simple transformation between them exists under the additional assumption that investors in the aggregate can be
 characterized as having a constant measure of risk-aversion, which we write as $\kappa$. Pronounced ``kappa", it's
 a single number, which I have estimated from historical SPX returns as $\kappa = 3 \pm 0.5$. More carefully, it's
 called the Coefficient of Relative Risk Aversion and has been heavily studied.\footnote{My estimates are in line with one of the
 most classical of these studies \cite{friend:1975}. See \cite{lewis:2019} for further commentary.} Indeed, to get from
$q$ to $p$, one just applies a simple exponential transformation

\be    p_{\bar{X}_T}(x) = \frac{\e^{\kappa x} \, q_{\bar{X}_T}(x)}{\int \e^{\kappa x} q_{\bar{X}_T}(x)  \, dx}.  \label{eq:measurechange} \eb 
 OK -- so how do we find $q_{\bar{X}_T}(x)$? My approach estimates $q$ by parameterizing
 it as a \emph{Gaussian mixture model}. Specifically, I take
  \be  q_{X_T}(x) = \sum_{i=1}^N w_i \,
 \frac{\e^{-(x - \mu_i \tau)^2/(2 \sigma_i^2 \tau)}}{\sqrt{2 \pi \sigma_i^2 \tau}}, \label{eq:GMM}
 \eb
 where $\tau = T-t$, and $N$ is  a small integer (5 in my fits). The fitted parameters
 are $N$ positive weights, $\{w_i\}$, and  $2N$ drifts and volatilities, $\{\mu_i,\sigma_i\}$.
 After a normalization and martingale condition, this leaves $3 N - 2$ free parameters at
 \emph{each} $(t,T)$ pair associated to a trade date and an option expiration. Free parameters are adjusted to fit option quotes:
  minimizing an objective function given in \cite{lewis:2019}. Finally, after algebra, now find -- on an annualized percent basis:
 
 \begin{empheq}[box=\fbox]{align}
 \mbox{ERP}_{t,T}^{(ann\%)}(\kappa) &= \frac{100}{T-t} \times
 \left\{  \left( \e^{\delta_{t,T} \tau} \sum_{i=1}^N \tilde{w}_i \, \e^{\alpha_i + (\kappa + \frac{1}{2}) v_i} \right) - 
 \e^{r_{t,T} \tau} \right\}, 
 \label{eq:ERPfinal}  \\
 & \mbox{using} \,\, \tau = T-t, \quad \alpha_i = \mu_i \tau, \,\, v_i = \sigma_i^2 \tau,  \nonumber \\      
 & \quad \,\, \gamma_i = \kappa \, \alpha_i  + \Smallfrac{1}{2} \kappa^2 v_i, \,\, \mbox{and}
 \,\, \tilde{w}_i = w_i \e^{\gamma_i}/\sum_{i=1}^N w_i \e^{\gamma_i}. \nonumber
 \end{empheq}  
 New parameters which have just appeared are $r_{t,T}$ and $\delta_{t,T}$: the cost-of-carry parameters. They correspond to the continuously
 compounded riskless rate and dividend yield associated to option expiration $T$. For example, if $C_{t,T}$ and $P_{t,T}$
 denote call and put prices with strike $K$, we have the model-independent, put-call parity relation:
 
 \be C_{t,T} - P_{t,T} = S_t \, \e^{-\delta_{t,T} \tau} - K \, \e^{-r_{t,T} \tau}
 = \e^{-r_{t,T} \tau} (F_{t,T} - K),   \label{eq:putcallparity} \eb
 where $F_{t,T}$ is an (option-implied) forward price. 
 
 Once all the parameters are estimated, we evaluate (\ref{eq:ERPfinal}) with $\kappa = 3$
 to get the mid-point estimates (dotted lines) for all the ERP charts in Sec. \ref{sec:timelines}. Similarly, we
 use $\kappa = 2.5$ and $\kappa = 3.5$ to estimate the lower and upper bounds to the uncertainty intervals (gray).  
 
 Appendix 2 discusses the computational details that either \emph{differ} from the discussion in \cite{lewis:2019} or
 would be unresolved if you simply turned to that reference.

\newpage

\section{Timelines with Equity Risk Premium Term Structures} \label{sec:timelines}

\subsection{Early days} 
\begin{figure}[h] 
	\caption{{\bf{Timeline 1}}} 
	\vspace{10pt}
	\begin{center}
		\includegraphics[width=0.9\textwidth]{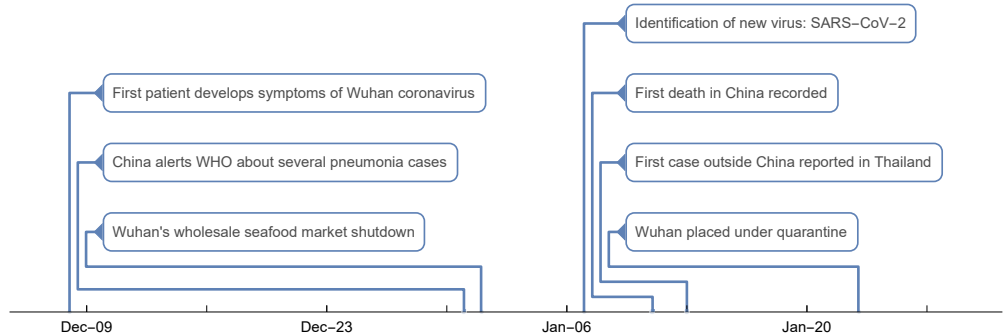}
	\end{center}		
	\label{fig:TimeLine1}
\end{figure}

Fig. \ref{fig:TimeLine1} shows some early events in the development of the 
pandemic.\footnote{Timeline events are drawn from the World Economic Forum (\url{https://www.weforum.org}), 
	the Wall Street Journal of Mar 21-22 2020, Zack's Equity Research (Stock market
	news), and misc. online news sources.} 

\begin{figure}[h] 
	\caption{{\bf{US ERP term structure}}} 
	\vspace{10pt}
	\begin{center}
		\includegraphics[width=0.8\textwidth]{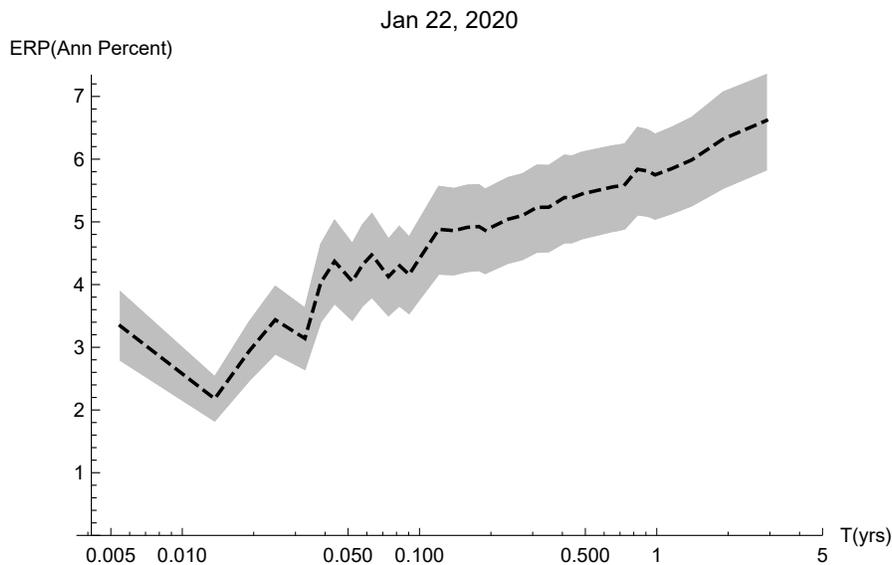}
	\end{center}		
	\label{fig:ERPts012220}
\end{figure}

\Pbreak
Wuhan, China was placed under quarantine on Jan 23, 2020, with rail and services suspended.  
Two days earlier, on Jan 21, the first US case was identified in Washington state  -- a man in his 30's who
had returned from a trip to Wuhan. Fig \ref{fig:ERPts012220} shows the estimated ERP term structure. It's 
within typical long-run ERP estimates of 3-6\% per year: the US equity market was not concerned.

\newpage

\subsection{First death in Europe} 
\begin{figure}[h] 
	\caption{{\bf{Timeline 2}}} 
	\vspace{10pt}
	\begin{center}
		\includegraphics[width=0.9\textwidth]{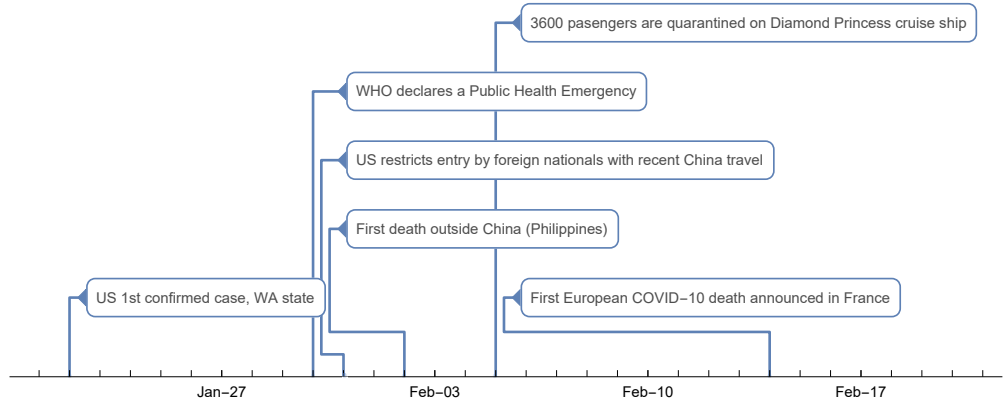}
	\end{center}		
	\label{fig:TimeLine2}
\end{figure}

Fig. \ref{fig:TimeLine2} shows some next events in the development of the pandemic.

\begin{figure}[h] 
	\caption{{\bf{US ERP term structure}}} 
	\vspace{10pt}
	\begin{center}
		\includegraphics[width=0.8\textwidth]{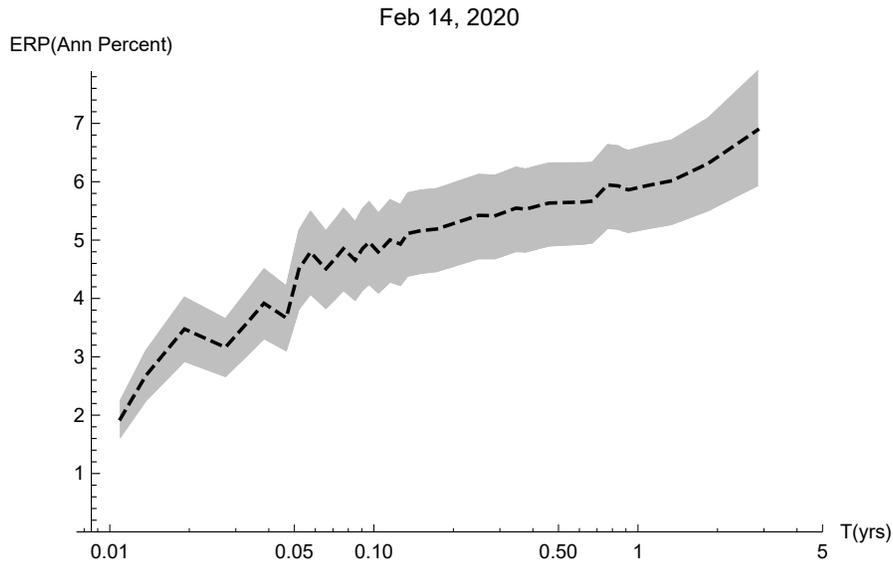}
	\end{center}		
	\label{fig:ERPts021420}
\end{figure}

\Pbreak
Feb 14, 2020 marks the end of this segment with the announcement of the first European COVID-19 death, in France. Fig \ref{fig:ERPts021420} shows the estimated ERP term structure. I think it's fair to say the US equity market remained  
in ``business as usual" mode.

\newpage

\subsection{Italy starts lockdowns} 
\begin{figure}[h] 
	\caption{{\bf{Timeline 3}}} 
	\vspace{10pt}
	\begin{center}
		\includegraphics[width=0.9\textwidth]{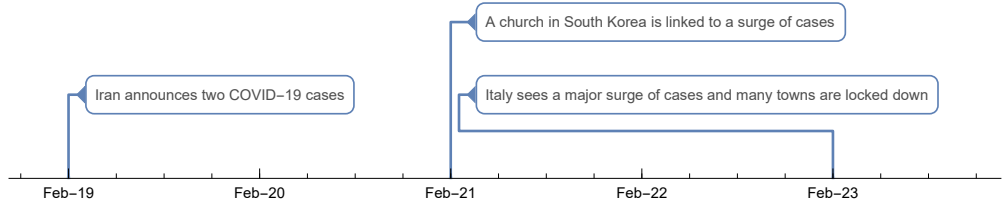}
	\end{center}		
	\label{fig:TimeLine3}
\end{figure}

Fig. \ref{fig:TimeLine3} shows some next events in the development of the pandemic.

\begin{figure}[h] 
	\caption{{\bf{US ERP term structure}}} 
	\vspace{10pt}
	\begin{center}
		\includegraphics[width=0.8\textwidth]{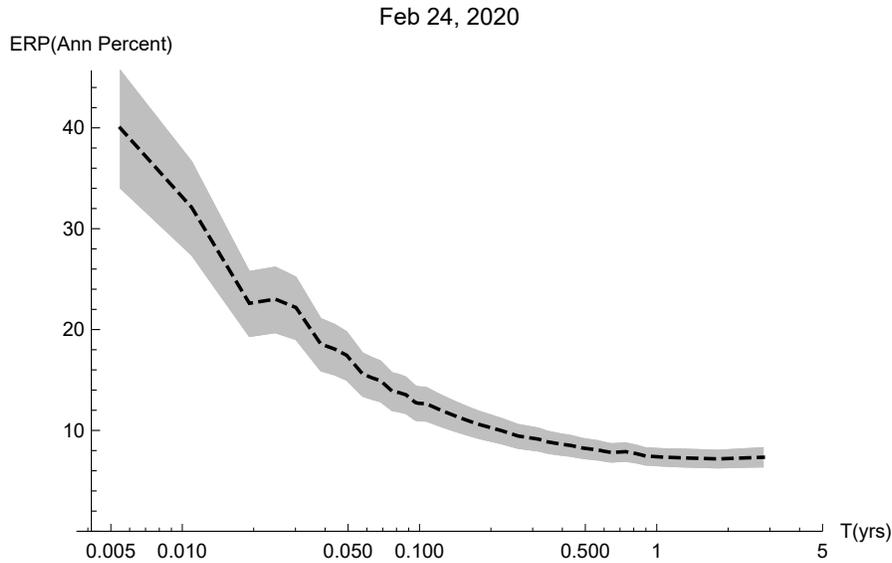}
	\end{center}		
	\label{fig:ERPts022420}
\end{figure}

\Pbreak
Feb 23, 2020 marks the end of this segment with start of lock-downs in Italy. As seen in Fig. \ref{fig:ERPts022420},
the ERP chart for the next day, the market is now definitely paying attention.
The S\&P500 has fallen 4.7\% from its record peak of 3386.15 on Feb 19, and the CBOE's VIX index has risen to 25.03. Fig \ref{fig:ERPts022420} shows the estimated ERP term structure. It's become strongly inverted with the short-term ERP (the ``required return") rising to about 40\%. Note the long end of 
the curve, representing the Dec 21, 2022 maturity -- about 2.8 years away. It's around 8\%, slightly higher but not too far from the longer-term value in the previous plots.

\newpage

\subsection{Early March -- the Fed acts} 
\begin{figure}[h] 
	\caption{{\bf{Timeline 4}}} 
	\vspace{10pt}
	\begin{center}
		\includegraphics[width=0.9\textwidth]{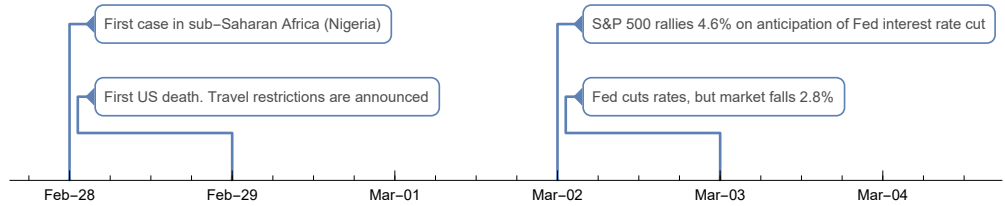}
	\end{center}		
	\label{fig:TimeLine4}
\end{figure}

Fig. \ref{fig:TimeLine4} shows some events from late February and early March.

\Pbreak
Early March begins a Federal Reserve monetary policy response. On Mar 2, 2020 the market rallies by almost 5\% -- likely anticipating  an interest rate cut. The VIX index (which closed at 40.1 on Feb 28), correspondingly eased to 33.4 on Mar 2. Indeed, the Fed announces a rate cut on Mar 3, the first unscheduled, emergency rate cut since the 2008-2009 Financial Crisis.
However, the market fell and the VIX climbed back to 36.8. 

\begin{figure}[h] 
	\caption{{\bf{US ERP term structure}}} 
	\vspace{10pt}
	\begin{center}
		\includegraphics[width=0.8\textwidth]{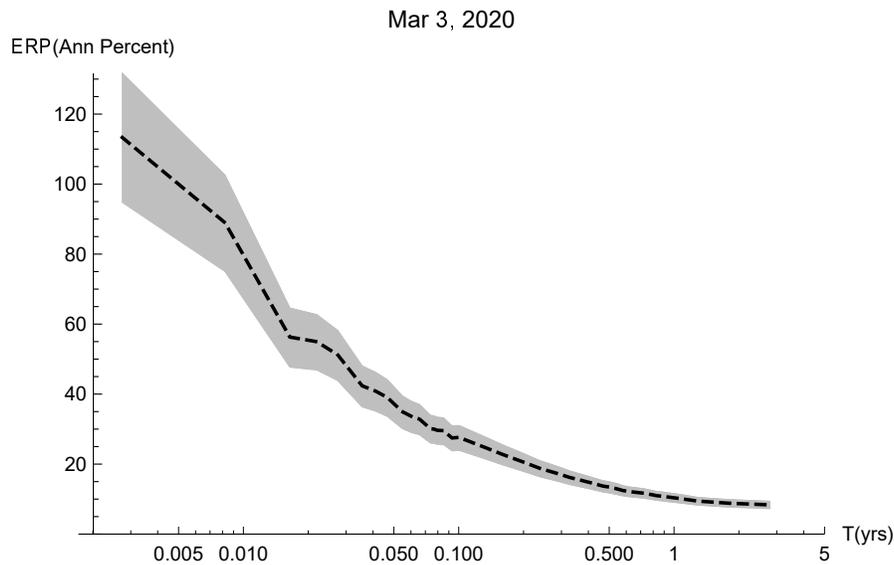}
	\end{center}		
	\label{fig:ERPts030320}
\end{figure}

\begin{figure}[h] 
	\caption{{\bf{VIX Index: longer run and latest one year}}} 
	\vspace{10pt}
	\begin{center}
		\includegraphics[width=0.9\textwidth]{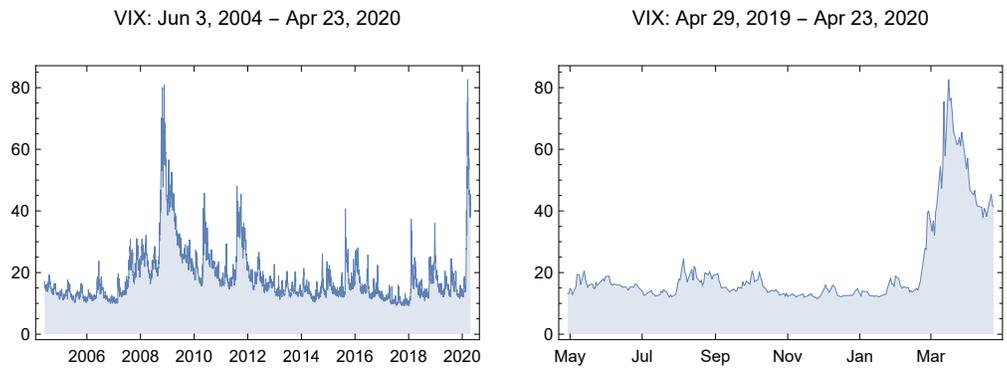}
	\end{center}		
	\label{fig:vixplots}
\end{figure}

In general, VIX's above 40 are a sign of a very high level of systematic market stress (see Fig. \ref{fig:vixplots}). The associated ERP term structure (Fig. \ref{fig:ERPts030320}) significantly steepens from our last plot, showing an estimated short-term required return of 100-130 percent per annum.

\newpage

\subsection{Mid March I -- time to panic} 
\begin{figure}[h] 
	\caption{{\bf{Timeline 5}}} 
	\vspace{10pt}
	\begin{center}
		\includegraphics[width=0.9\textwidth]{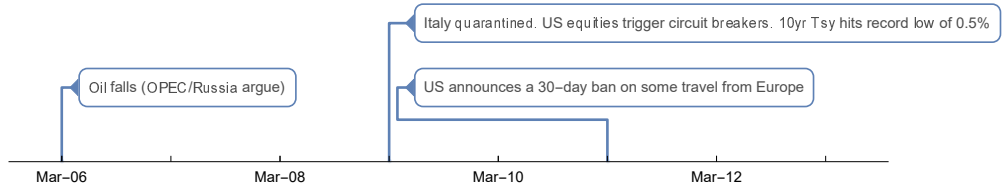}
	\end{center}		
	\label{fig:TimeLine5}
\end{figure}

Fig. \ref{fig:TimeLine5} shows some events from mid-March.

\Pbreak
The week of Mar 9-13 is very ugly on many fronts. COVID-19 cases are rising exponentially in Europe and the US. In the
absence of mitigation, there are predictions of millions of deaths to occur in the US before so-called `herd immunity' is achieved. A vaccine is predicted to be at least 18 months away, and not certain even then. 

On Thurs Mar 12, 2020, the VIX index closes at 75.5\% and the short-term (1-day) annualized
ERP reaches 500-600\% (see Fig. \ref{fig:ERPts030320}). Although hard to discern in the chart,
the ERP over the longest term ($2 \Smallfrac{3}{4}$ years) has risen to approximately 17\% annualized. Credit markets are highly stressed.

\newpage

\subsection{Mid March II -- lockdowns, the IHME becomes influential} 
\begin{figure}[h] 
	\caption{{\bf{Timeline 6}}} 
	\vspace{10pt}
	\begin{center}
		\includegraphics[width=0.9\textwidth]{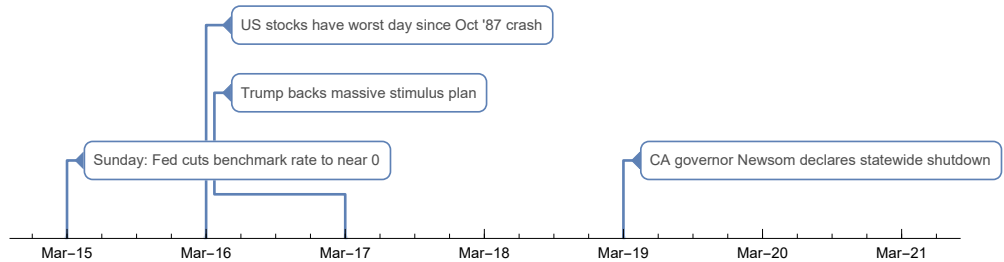}
	\end{center}		
	\label{fig:TimeLine6}
\end{figure}

Fig. \ref{fig:TimeLine6} shows some events from mid-March.

\begin{figure}[h] 
	\caption{{\bf{US ERP term structure}}} 
	\vspace{10pt}
	\begin{center}
		\includegraphics[width=0.8\textwidth]{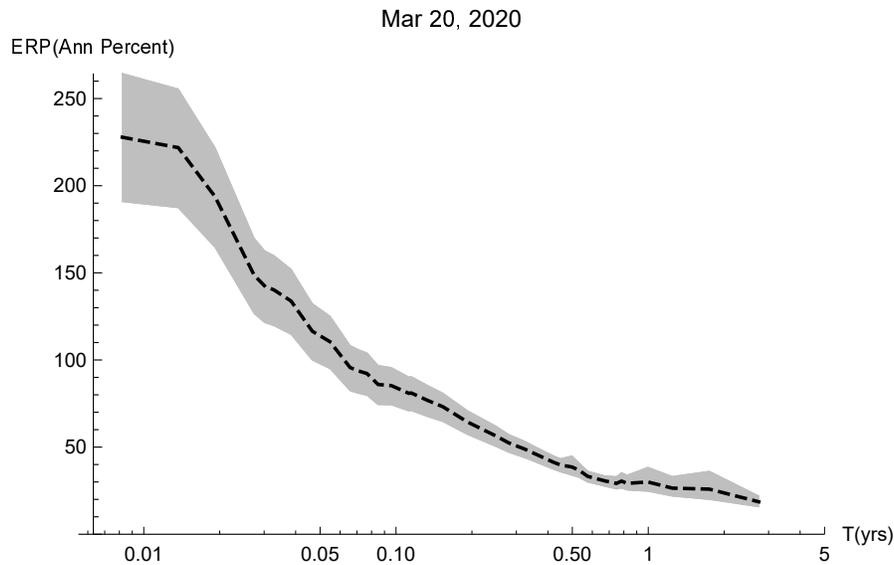}
	\end{center}		
	\label{fig:ERPts032020}
\end{figure}

\Pbreak
By the end of Mar 16-20 week, stresses begin to ease somewhat: see  Fig. \ref{fig:ERPts032020}. 
The short-term ERP estimate ended the week at 200-250\%, the lowest of the week. VIX ended at 66.

What prompted the ease? Prospective stimulus likely helped. Also, the U. Washington's Institute for Health Metrics and Evaluation (IHME) was gaining influence.\footnote{See \url{https://covid19.healthdata.org/united-states-of-america}} Their earliest predictions (Mar 25), based upon curve fitting to the Wuhan experience, suggested cumulative US deaths to total 38,000-162,000 through Aug 2020 -- premised on lock-downs. These estimates were significantly lower than the 
previous `millions' of others (without mitigation), and not terribly disproportionate to the annual mortality
from the flu. Indeed, as time passed, the IHME US death estimates were tightened and lowered: 
48,000-123,000 at this writing (late April 2020), with 54,000 actual deaths to date.

\newpage      

\subsection{Late March to mid-April -- signs of optimism} 
\begin{figure}[h] 
	\caption{{\bf{Timeline 7}}} 
	\vspace{10pt}
	\begin{center}
		\includegraphics[width=0.9\textwidth]{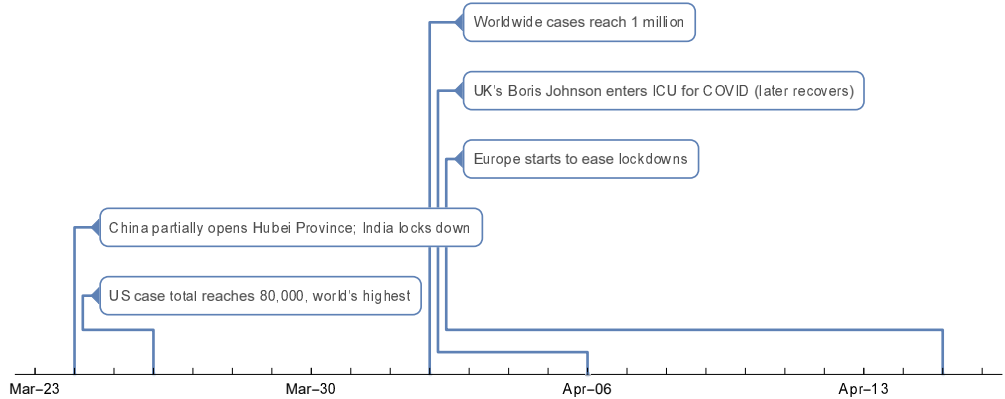}
	\end{center}		
	\label{fig:TimeLine7}
\end{figure}

Fig. \ref{fig:TimeLine7} shows some events through April 15, 2020.

\Pbreak
April 15, 2020 marks the end of our ERP study. The last timeline events reflect increased optimism that
(at least the initial phase of) the pandemic has plateaued or peaked in many areas of the world.  VIX ended at 40.8, and the
short-term ERP has fallen to the 50-60\% range. The long end of the ERP curve remains quite elevated
at 16.7\%.

\begin{figure}[h] 
	\caption{{\bf{US ERP term structure}}} 
	\vspace{10pt}
	\begin{center}
		\includegraphics[width=0.8\textwidth]{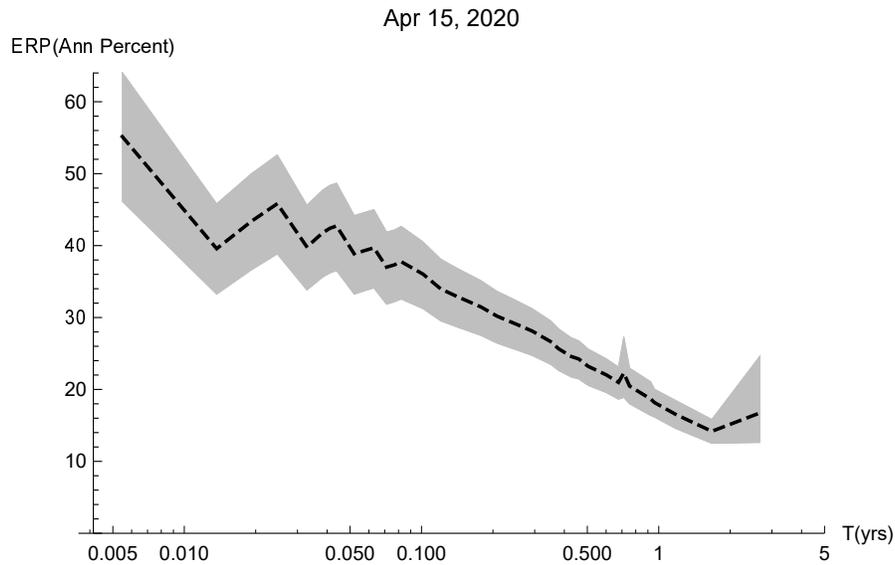}
	\end{center}		
	\label{fig:ERPts041520}
\end{figure} 

\newpage

\section{Outlook and Future Work}
  The outlook for the course of the pandemic and the economy at this writing is encouraging. For example, in New York state, the
  hardest hit US state, hospitalizations are down significantly from a month ago. Encouraged by that, Governor Andrew Cuomo, says he will extend the ‘PAUSE’ regulations in many parts of the state, but some less-affected regions can reopen on May 15. 
  I hope to see the same in my state, California. It's clear
  that reopening will be done carefully everywhere, with social distancing and protective measures an ongoing recommended
  part of life for many months to come. 
  
  There are many to-be-answered questions: what exactly is the mortality rate, how many have been infected, are infected
  but asymptomatic or recovered people now immune, etc? Several recent studies suggest that the mortality rate is much lower
  than many original estimates.  
  
  In terms of my financial analysis here, there are also some unanswered questions. For example, what exactly is the
  risk-return trade-off here? This can be answered by computing the (annualized) variance rate 
  $\sigma^2_{t,T}$ associated to the inferred real-world $p$-distributions, and plotting $\mbox{ERP}_{t,T}$ vs.  $\sigma^2_{t,T}$.  	
   Another project on my ``to-do" list is to organize the ERP's for \emph{standardized} maturities, say
  3 days, 1 week, 1 month, and so on. Finally, I would like to continue to update the results as time progresses. As I work through these projects, I'll update this preprint.

\newpage

\section{Appendix 1 -- Basic reference charts} 

\newpage

\newpage

\section{Appendix 2 -- More computational details} 
  Here, I discuss some differences from my previous study: \cite{lewis:2019}.

\pbold{Data.} As in my previous study, option quotes were sourced from the CBOE’s LiveVol service: ``End-of-Day Option Quotes with Calcs". These quotes are recorded at 15:45 New York time, 15 minutes from the close of the regular session. The CBOE advertises
them as a ``more accurate snapshot of market liquidity than the end of day market".

With my previous study data, every non-zero bid option was accompanied by a larger non-zero ask. In my data for this study, which was sampled almost completely for all
SPX trade dates and quotes from Jan 2, 2020 through April 15, 2020, there were a few exceptions to this rule. For example,
on March 16, 2020 there were some quotes for the Dec 16, 2022 expiration showing a bid$>0$ but an ask=0. I sent a query
to the CBOE and a staff person explained that, first an ask=0 value should be interpreted as ``no ask present". And similarly
on the bid side. He also explained that this can occur as ``Liquidity providers temporarily vacate the quoting environment
to protect themselves while they re-evaluate their assumptions". There were very few instances of this, and my
procedures seemed to suffer no ill-effects by simply ignoring these strikes. Nevertheless, I thought it was unusual and worth
reporting.  

\pbold{Dual expirations.} In my previous study, I included both AM and PM options on the Fridays where these both occurred.
These were treated as distinct expirations because times were measured to 15 min accuracy.  
Here, for simplicity, for such dual expirations, only the PM options were included. However, certainly the AM options were
included here for any expiration when those were the sole options expiring. Also for simplicity, all times $T$ here (in years),
were simply measured as T=(days)/365, where days was the integer number of days from the trade date to the expiration date. 

\pbold{Cost-of-carry methodology.} I adopted the put-call parity regression method of the previous study. One change was that
I only included 50\% of the put-call pairs: those closest to the money. For very short-dated expirations, while the
regression method produces very plausible forward prices (which are key), the inferred interest rate and dividend yield are rather erratic, often negative. This erratic effect was lessoned by the 50\% inclusion (the previous study using 100\%). The previous study showed that
two different cost-of-carry methods will produce almost identical ERP's as long as the inferred forward prices are similar.  

\pbold{Simplified rules.} In my previous study, besides my nominal
objective function, I also adopted a secondary objective related to certain {\MM{OutStats}}, which are explained there.
I sought to achieve my secondary objective by switching the number of Gaussian components from N=4 to N=5, or making other
adjustments such as introducing a minimum bid on put quotes higher than the $0.05$ non-zero minimum characteristic of SPX
options. Here, for both simplicity and efficiency, I fixed on N=5 Gaussian components for the entire study and always included
every out-of-the-money option quote with both a non-zero bid and non-zero ask. I declared these to be ``simplified rules",
and confirmed that all the ERP's of my previous study were reproduced to 3 good digits under my simplified rules. 

The {\MM{OutStats}} here were not as good as in my previous study, but generally I found my ERP's to be quite robust to
my choices for various optimizer parameters. Optimizer parameters consisted of the PrecisionGoal (PG=4),
the maximum number of optimizer steps (MAXSTEPS=600), and a scaling parameter ({\MM{sigMULT}=1.2}). This last one
fixed the upper limit of the allowed fitted volatility to be {\MM{sigMULT}} $\times$ {\MM{IVMAX}}, where the second term was the highest implied volatility observed at that expiration. On a few expirations, if the  {\MM{OutStats}} looked
particularly poor, I would try a rerun with MAXSTEPS=1000, or PG=5 or sigMULT=1.4. Typically, the fit would improve,
but with the ERP either unchanged or only changing slightly. That's what I mean by `robust'. 
Overall, I think my estimates are good to three digits.

\end{document}